\documentclass[preprint,showpacs,aps,amsmath]{revtex4}
\usepackage{graphicx}
\usepackage{bm}

\begin{document}
\preprint{}
\title{Transcranial stimulability of phosphenes \\ by long lightning
  electromagnetic pulses}
\author{J. Peer and A. Kendl\footnote{Corresponding author: alexander.kendl@uibk.ac.at}}
\affiliation{Institut f\"ur Ionenphysik und Angewandte Physik, Universit\"at
  Innsbruck, Austria} 

\begin{abstract}
The electromagnetic pulses of rare long (order of seconds) repetitive
lightning discharges near strike point (order of 100~m)
are analyzed and compared to magnetic fields applied in standard clinical
transcranial magnetic stimulation (TMS) practice. 
It is shown that the time-varying lightning magnetic fields and locally induced
potentials are in the same order of magnitude and frequency as those established in
TMS experiments to study stimulated perception phenomena, like magnetophosphenes.
Lightning electromagnetic pulse induced transcranial magnetic stimulation of
phosphenes in the visual cortex is concluded to be a plausible interpretation
of a large class of reports on luminous perceptions during thunderstorms.\\
$[$Physics Letters A, Volume 374, Issue 29, 28 June 2010, Pages 2932-2935$]$


{\begin{center} \large APPENDIX: Erratum and Addendum \end{center}}

The comparison of electric fields transcranially induced by lightning
discharges and by TMS brain stimulators via $\vec{E}=- \partial_t \vec{A}$ is
shown to be inappropriate. Corrected results with respect to evaluation of
phosphene stimulability are presented.
For average lightning parameters the correct induced electric fields
appear more than an order of magnitude smaller.
For typical ranges of stronger than average lightning currents,
electric fields above the threshold for cortical phosphene stimulation can be
induced only for short distances (order of meters), or in medium distances
(order of 50~m) only for pulses shorter than established axon excitation periods. 
Stimulation of retinal phosphene perception has much lower threshold and
appears most probable for lightning electromagnetic fields.\\
$[$Physics Letters A, Volume 374, Issue 47, 2010, Pages 4797-4799$]$

\end{abstract}

\maketitle

{\noindent \bf Introduction}

\medskip

Transcranial magnetic stimulation (TMS) of neural activity in the human brain
has developed into an established method for neurophysical medical diagnosis
and psychiatric treatment~\cite{Hallet00,Walsh00}. In particular, stimulation of the visual
cortex by pulsed magnetic fields directed at suitable positions towards the
head has been reported to invoke phosphenes in probands, which are perceived
as luminous shapes within the visual field~\cite{Marg91}. Here we show that the
near-field electromagnetic pulses of natural rare long (1-2~s) repetitive
lightning strokes can be expected to lead to neural induction currents above
threshold values in the same order of magnitude regarding frequency, duration
and strength of stimulation as used in medical TMS. For a small fraction of
lightning flashes a near observer (ca. 20-200~m) should experience repetitive
stimulation of perception activity similar to clinical TMS effects.  We
conclude evidence for a plausible interpretation of a large class of reports
on luminous phenomena during thunderstorms as lightning electromagnetic pulse
induced transcranial magnetic stimulation of phosphenes in the human brain. 
An observer is likely to classify such an experience under the preconcepted
collective term of "ball lightning". 

\bigskip

{\noindent \bf Motivation: the phosphene interpretation of ``ball lightning'' reports}

\medskip

According to a comprehensive review by Stenhoff~\cite{Stenhoff99}, 
``ball lightning'' (BL) has been reported in the open air, indoors, and within
aircraft. 
Around one third of BL events may be attributed to observations of
stationary corona discharges in strong thunderstorm electric
fields~\cite{Stenhoff99}. 
The majority of observations which have been analyzed in different surveys (cited
{\it ibidem}) reported BL to be directly succeeding a cloud-to-ground lightning flash. 
Some hypothetical scenarios for BL-like dust-gas fireballs appearing in very specific
environmental situations after a stroke in sand or water have been suggested as a
possible explanation \cite{Abrahamson00,Paiva07,Versteegh08}. 

We here propose that a large class of reports (about the half) characterizing BL
as luminous roundish objects arising in coincidence with lightning flashes and
appearing to move slowly at eye level of an observer for a few seconds (often
accompanied by whitish noises and smells) can be interpreted as magnetic
phosphenes. 

The phosphene interpretation of ``ball lightning'' has been proposed earlier by
J. Swithenbank (reported in Ref.~\cite{Stenhoff99}) after personal BL observation, and was
discussed (and first brought in context with TMS) in a skeptical review of BL
theories \cite{Kendl01}. Other authors have in Ref.~\cite{Keul08} cursorily dismissed the
phosphene hypothesis with an erroneous argument considering only magnetic
field strengths (and not indeed their time derivative) and only the short
pulses of single stroke flashes. Recently, Cooray and Cooray \cite{Cooray08} have 
presented a somewhat related hypothesis of BL-like visual perceptions to be
possibly caused indrectly by epileptic seizures that may also be triggered by
lightning electromagnetic pulses. A comprehensive review of other (more or
less plausible) BL theories is given in Ref.~\cite{Stenhoff99}.

In the following we show that the electric fields induced by nearby long
repetitive lightning strokes are indeed sufficient to evoke the perception of
magnetophosphenes in the occipital cortex.

\bigskip

{\noindent \bf Magnetophosphenes: visual perception by induction}

\medskip

The normal process of visual perception comprises the conversion of optical stimuli
into electric signals by photoreceptors in the retina, and subsequent
propagation of sensor potentials to the visual cortex in the occipital brain
by neuron networks.
Transmission of stimuli occurs in form of action potentials caused by
processes opening and closing selective ion channels  in the cell
membranes. Action potentials form irrevocably if the depolarization of a cell
membrane due to external stimuli exceeds a threshold value of $U_{thr} \sim
20$~mV above the resting potential (-50~mV $> U_{rest} >$ -70~mV). The intensity of a 
stimulus is encoded by the frequency of subsequent action potentials~\cite{Kammer05}.  

Magnetic phosphenes are visual perceptions caused by time varying magnetic
fields $\bm B (\bm x, t)$, described by the vector potential $\bm A(\bm x,t)$ from $\bm B
=\nabla \times \bm A$, that induce sufficiently strong electric fields 
$\bm E_{ind} (\bm x,t) = - {\partial_t \bm A (\bm x, t)}$
to cause a local potential (determinded via  $\bm E_{ind} (\bm x,t) = - \bm
\nabla U_{ind} (\bm x,t)$) on the membrane exceeding $U_{ind} > U_{thr}$.
These change the membrane potential and trigger an action potential either in the
retina, in transmitting neurons, or directly in neurons of the visual cortex.  
The resulting visual perception is termed retinal phosphene or cortical phosphene,
respectively, according to the location of the stimulus at the retina or in the cortex.

\bigskip

{\noindent \bf Cortical phosphenes induced by transcranial magnetic stimulation}

\medskip

Transcranial magnetic stimulation (TMS) is a method for noninvasive selective
magnetic stimulation of local brain areas~\cite{Hallet00,Walsh00}. 
Perceptible stimulation can be achieved by application of either single
magnetic pulses or by repetitive pulses (rTMS) through stimulation coils
placed on the outside of the head. 
Typical duration of a single neural TMS pulse is in the order of
250-450$\mu$s, and typical repetitive pulse frequencies are in the range of
1-50~Hz. 
The transient magnetic field induces a local electric field inside the brain
which can form an action potential in the stimulated area if $U_{ind} > U_{thr}$. 

Cortical phosphenes, which are perceived as luminous shapes within the visual
field, are reported when the TM stimulus is applied to the area of the visual
cortex and the local induced field amplitude exceeds values in the range of
20-50~V/m, with varying thresholds in different subjects~\cite{Marg91}. Phosphenes are
perceived in various shapes (ovals, bubbles, lines, patches) within the visual
field, mostly appearing white, gray or in unsaturated colours~\cite{Kammer05b}. The
duration of perception follows the duration of the single pulses or the whole
repetitive cycle respectively. Phosphenes appear moving when the stimulation
coil is shifted or the fixation site is changed. Impressions appear stronger
and brighter with increasing stimulus strength~\cite{Kammer99}.  

Retinal phosphenes have even lower threshold values than their cortical
counterparts~\cite{Litvak02,Kavet08}. Motivated by the availability of many well
documented clinical TMS studies on cortical phosphenes, and by the established
specifications of TMS induction coils, we restrict to those in the following 
comparison with lightning electromagnetic pulses (LEMPs). 

\bigskip

{\noindent \bf Repetitive LEMPs and TMS}

\medskip

Phosphenes in clinical TMS are reported to occur only during the actual
duration of stimulation (without significantly longer lasting after effects). 
A perception caused by LEMPs can therefore be duely expected for duration at
least comparable to or longer than typical TM stimulation experiment times of
250-450~$\mu$s.  

Negative (CG-) downward discharges occur in 90\% of cloud-to-ground lightning.
Typical CG- discharges begin with an electric stepped leader breakdown and a
first return stroke, and are in most cases followed by multiple subsequent
strokes, which are each initiated by a dart leader pulse through the pre-established channel. 
Single stroke CG- flashes have a typical duration of several hundred microseconds.
Positive cloud-to-ground flashes (CG+) have rarer occurence and are usually
limited to a single stroke, but may occur with higher continuing currents for
longer discharge times up to $0.1$~s \cite{Rakov06}.

Stimulation by single stroke CG- or CG+ discharges may, as a consequence, 
cause brief phosphene perceptions (if the stimulus strength is above threshold),
but is not able to explain reported BL durations in the order of seconds.

Long CG- flashes consisting of repetitive strokes occur at stroke intervals
between 4-500~ms with a mean value of 50~ms~\cite{Rakov06}, which in fact are
exactly compatible to standard rTMS frequencies in the range of 1-50~Hz.  
Phosphene perception by clinical rTMS has been reported for 3-5 successive pulses or
more. The average number (multiplicity) of lightning strokes per flash is also
between $n=$2 and 5, but more than 20 strokes per flash with a total duration
up to two seconds have been observed in detection networks~\cite{Rakov03}. 
Further subsequent strokes (possibly up to more than 40) with decreasing
amplitudes often fail to enter the statistics by not exceeding the threshold
of remotely distributed detectors.  

Although the electromagnetic pulses of the stepped leader and first return stroke could
lead to induced fields above the phosphene threshold, these are of minor
importance for the long term field evolution of high multiplicity flashes. 
The further discussion can be limited to the effects of following dart leaders and
subsequent return strokes. Repetitive stimulation by these multiple return strokes of
$n>20$ can occur with durations $t > 20 \cdot 50$~ms in the order of several seconds.

\bigskip

{\noindent \bf Calculation of lightning electromagnetic fields}

\medskip

Now we address the question if natural repetitive cloud-to-ground LEMPs
generated by nearby strokes are able to 
transcranially induce electric fields comparable to those generated by
clinical TMS (of around 20-50~V/m), and thus sufficiently strong to
stimulate similar sensory perceptions. 

For this purpose we have calculated the near electromagnetic fields of
lightning discharges for various types and parameters of naturally occuring
flashes. Previously published field calculations have mostly been restricted
either to far fields ($>$~km) relevant to lightning detection networks, or to
direct impacts relevant to engineering problem of lighting protection.  

The model and numerical methods of our near field LEMP calculations, including
the effects of channel tortuosity and arbitrary observer location, are based
on Refs.~\cite{Rakov06,Thottappillil97,Thottappillil94,Cooray04}. 
For details on the method and general results we refer to Ref.~\cite{Peer10}: 
Maxwell's equations are
integrated including retardation without scale approximations for given
lightning channel base currents to yield the electric field $\bm E(\bm x,t)$ and
electromagnetic vector potential $\bm A(\bm x,t)$ depending on time $t$ and location
$\bm x$. Induced electric fields at location $\bm x_o$ of a near observer (20-100~m
horizontal distance from impact, level to perfectly conducting ground) are
derived from the time derivative of the vector potential $\bm A(\bm x_o,t)$ for various
stroke types such as leader, return strokes and M-components. For simplicty,
cortical anisotropy and dielectric properties have been neglected in this
work.

\bigskip

{\noindent \bf Results: stimulation induced by successive return strokes}

\medskip

We first consider straight vertical lightning channels using a leader model with a
typical value for the homogeneous charge distribution of $q = 0.14$~mC/m
\cite{Rakov06}, and a current generation type model for the return stroke~\cite{Cooray04}.

Our numerical calculations on subsequent mutiple CG- dart leaders and return
strokes show that in distances of the order of 20-100~m only the latter can
induce above electric fields long enough to envoke perception: 
induced electric fields of dart leaders can in fact reach  $E_{ind} > 20$~V/m
above threshold, but the short dart leader pulse period of 2-3~$\mu$s (compared to TMS
pulses of several 100~$\mu$s) may prohibit actual cognitive perception. 

Return strokes are characterized by a fast rising phase and a slower decline
phase of the nearby local magnetic field strength. 
The calculated pulse shapes of transcranially induced electric fields begin with a strong
field peak in the order of kilovolts per meter and duration of microseconds caused by
the large time derivative of the magnetic field in the short rise phase. 

The action of this initial peak is difficult to predict due to the lack of
comparably sharp field pulses in clinical brain stimulation.
Assuming that the cell membrane of a single axon can be modelled as an RC circuit
(i.e. a capacitance with a parallel connected leakage resistance)
characterised by the cortical time constant of 150~$\mu$s~\cite{Kammer05,Barker91}, 
we can roughly estimate the resulting change in the membrane potential: 
The time dependent capacitor charging voltage  of an RC circuit is given by
$U(t)=U_0\left(1-\exp(-t/\tau)\right)$ where $U_0$  is the applied voltage and
$\tau$ is the time constant (i.e. the time taken by $U(t)$ to increase to 63\% of $U_0$). 
Considering that a TMS induced electric field pulse of 20~V/m and 300~$\mu$s
is able to trigger an action potential, we can deduce from the above equation
that the initial field peak of a lightning return stroke ($E_{ind} \approx
2$~kV/m, $t \approx 0.5\mu$s) leads to a membrane depolarisation which is
$\left(2000\left(1-\exp(-\frac{0.5}{150})\right)\right)
/\left(20\left(1-\exp(-\frac{300}{150})\right)\right) \approx 0.4$ times the  
depolarisation relative to the considered TMS pulse. 
Thus, an action potential could  already be caused by this strong initial field
rise phase peak of the return stroke

\begin{figure}
\includegraphics[width=8.0cm]{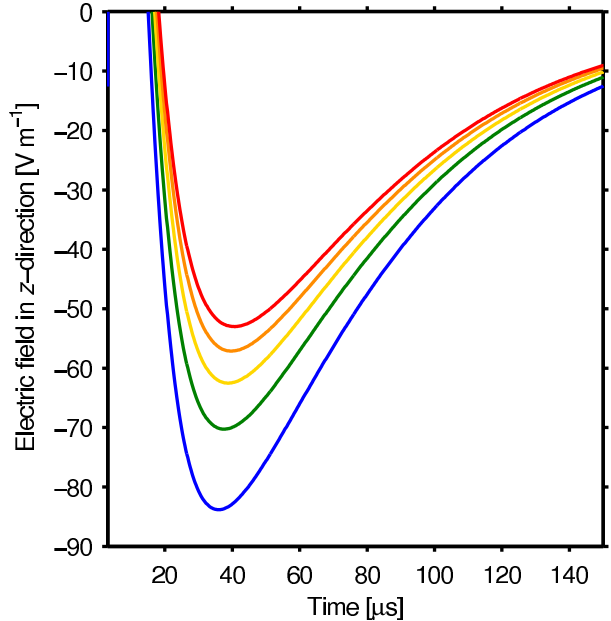}
\caption{\label{f:bfield} 
\sl Electric field transcranially induced at various observation points
(from bottom to top: 20 - 100m distance from strike point) by the time derivative of the lightning
magnetic field during the decline phase of one average negative cloud-to-ground
subsequent return stroke within a long high-multiplicity flash.}
\end{figure}

The following long decline phase of return strokes LEMPs in the order of 200~$\mu$s
has the most relevance for stimulation:
our calculations for average discharge parameters show that in this phase LEMP
induced potentials of the same order in amplitude and duration as rTMS pulses
(larger than 20-50~V/m) occur in a distance less than around 100~m from the lightning channel.
Results of the detailed simulations of this last phase of CG- return strokes are shown in
Fig.~\ref{f:bfield} for various observer distances and otherwise standard parameters.

High multiplicity lightning, which has similar pulse repetition frequency as
rTMS, can therefore be positively expected to stimulate cortical phosphenes
for as long as several seconds. 
The observation of magnetophosphenes is actually not restricted to distances below
100~m, but may be experienced up to 300~m from the impact point, as strokes
can occur with intensities (channel currents) up to 10 times larger than the
average values used in our calculations.

\bigskip

{\noindent \bf Conclusion: likelihood to experience magnetophosphenes during a thunderstorm}

\medskip

In summary, we have calculated and analyzed the electric fields induced by all
phases of near multiple lightning electromagnetic pulses, and have shown a
remarkable agreement with fields induced by repetitive transcranial magnetic
stimulation, which is known to cause phosphene perception in observers when
applied to the visual cortex.

The chance for transcranial stimulability of LEMP induced phosphenes can be
roughly estimated. Occurence of a repetitive stroke near to an observer ($ <
{\cal O}(200$~m)) is essential to achieve an above threshold induction 
potential. Noticeable perception of phosphenes very likely occurs only when
other sensory stimuli (or bodily injury of the observer) are not
dominant. Direct observation of the blinding light and deafeningly loud thunder of
lightning bolts may drown out phosphene perception. Magnetic fields of 
LEMPs are however able to penetrate walls and roofs, so that a direct line of
sight to the bolt is not necessary to experience phosphenes.

Long perception in the order of seconds can be expected for the more
rarely occuring repetitive strokes with multiplicity higher than 20,
which occur for 1-5\% of CG- strokes, although published statistics of such
events are scarce~\cite{Rakov06}. 
As a conservative estimate, roughly 1\% of (otherwise unharmed) close lightning
experiencers are likely to perceive transcranially induced above-threshold
cortical stimuli. The activation by (time varying) weakly damped penetrating
magnetic fields allows observation within closed buildings or
aircrafts. Broadband stimulation of other sensory activity (odours, sound) can
also be expected, but visual stimuli are usually dominantly perceived. 

An observer reporting this experience is likely to classify the event under the
preconcepted term of "ball lightning", which is used to subsume numerous
reports on luminous perceptions during thunderstorm activity~\cite{Stenhoff99}. 

Here we conclude evidence for interpretation of a large class of "ball
lightning" observations as magnetic phosphenes transcranially stimulated by
nearby long repetitive lightning strokes.

\bigskip

{\noindent \bf Acknowledgements}

\medskip

We thank Dr. Thomas Kammer (Head of the Laboratory for Transcranial Magnetic
Stimulation), Department of Psychiatry, University of Ulm (Germany),  for
valuable discussions, advice and careful reading of the manuscript. The
computational work has been funded by a junior research group grant
(``Nachwuchsf\"orderung") from the University of Innsbruck.

\newpage

{\begin{center} 
{\large \bf Erratum and addendum}\\
J. Peer$^{1}$, V. Cooray$^{2}$, G. Cooray$^{3}$, A. Kendl$^{1}$\\
\sl
1) Institute for Ion Physics and Applied Physics, University of
  Innsbruck, Austria\\
2) Division for Electricity, Department of Engineering Sciences, Uppsala
  University, Sweden\\
3) Department of Neurophysiology, Karolinska Institute, Sweden
\end{center}}

In Ref.~\cite{peer2010a} the electric fields $\vec E_\mathrm{ind}$ induced in the
head of a nearby observer by natural lightning discharges (LD) were compared to 
laboratory transcranial magnetic brain stimulation (BS) fields and effects. 
In this respect an inappropriate assumption has been applied, that both 
$\vec E_\mathrm{ind}^\mathrm{LD}$ and $\vec E_\mathrm{ind}^\mathrm{BS}$ could be calculated by 
\begin{equation}
  \vec{E}=-\partial_t \vec{A},
  \label{eqn:edta}
\end{equation}
which is valid if an electrostatic contribution $-\nabla \phi$ to the
right hand side due to space charge accumulation can be neglected. In the
following we show that this assumption is normally valid for BS but not for LD.

The vector potential ${\vec A}$ in the proximity of a straight vertical lightning
channel is also directed vertically and its magnitude is decreasing with distance.  
In the case of a circular TMS field coil ${\vec A}$ is again oriented like the
direction of the current flow, but here the current and therefore also 
${\vec A}$ form closed loops inside the head, which are approximately parallel
to the skull surface and do not necessarily cut through any surfaces. Hence there will be
no charge accumulation (and hence no buildup of an electrostatic potential
$\phi$), if the cortex is assumed to be an isotropic conducting medium. 

\begin{figure}[htb]
  \begin{center}
    \includegraphics[width=0.8\textwidth]{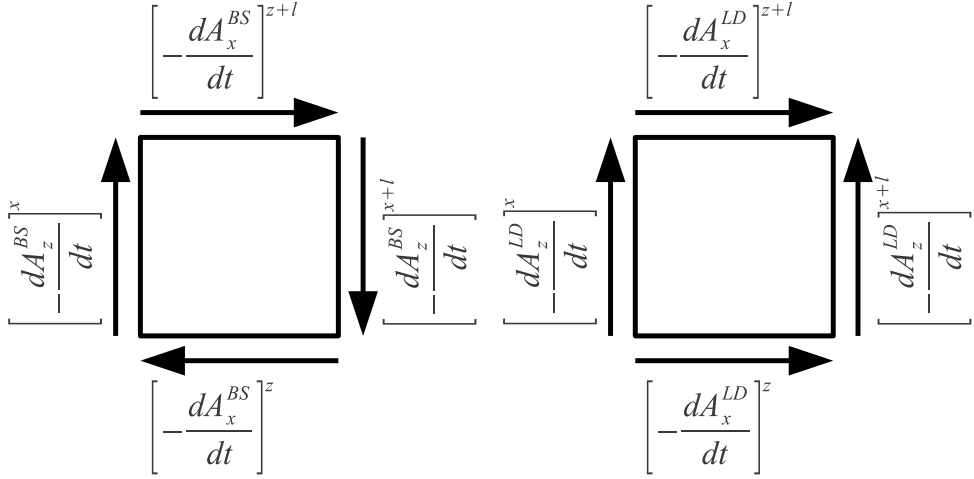}
  \end{center}
  \caption{\sl Electric fields around a quadratic loop due to the vector potential of a brain stimulation coil (left) and the vector potential of a lightning channel pointing in $z$-direction (right).}
  \label{fig:sketch}
\end{figure}

Fig.~\ref{fig:sketch} shows the direction of components of ${\vec E} = -
d{\vec A} /dt$ projected onto a "quadratic loop" inside the head. 
For clinical brain stimulation, the components form a closed loop ("BS", left
figure part), while for a lightning magnetic field there is a net contribution
from one corner to its opposite on the loop ("LD", right part). 

If, as it is the case for lightning fields, the vector potential does cut a
surface (of the cortex or the skull), across which there are two media of
different conductivity, there will be charge accumulation on the surface.  
This will cause a non-zero scalar potential $\phi$ which must be included
in calculating the total electric field. However, in the complex geometry of the
different conducting media in the head an exact calculation is a highly
nontrivial task. 

More generally, the electric field $\vec E(t)$ induced by a time varying
magnetic $\vec B(t)$ field is calculated from the Maxwell-Faraday equation 
$ \nabla \times \vec{E} = - \partial_t \vec{B} $
so that 
\begin{equation}
U_{\mathrm{ind}} = \oint \vec{E} \cdot \vec{dl} = - \partial_t \int \vec{B}
\cdot \vec{dS} = - \partial_t \psi_{}
  \label{eqn:far}
\end{equation}
corresponds to the voltage induced in a loop surrounding an area $S$,
enclosing the magnetic flux $\psi_{}$.
The average electric field along the loop can be calculated by
$\langle E \rangle = U_{\mathrm{ind}} / L = \partial_t \psi / L$ where $L=\int dl$.

In the literature concerning clinical BS (e.g. Ref.~\cite{salinas2007a}),
eq.~\eqref{eqn:edta} is used to compute $E_\mathrm{ind}^\mathrm{BS}$.
In Ref.~\cite{peer2010a} we therefore used expression~\eqref{eqn:edta} as a
reference quantity for the comparison of $E_\mathrm{ind}^\mathrm{LD}$ and
$E_\mathrm{ind}^\mathrm{BS}$. 
Eq.~\eqref{eqn:edta} indeed corresponds to $E_\mathrm{ind}^\mathrm{BS}$
when it is applied to brain stimulation coils. 
$E_\mathrm{ind}^\mathrm{LD}$, however, is different from eq.~\eqref{eqn:edta}
because of the different spatial variation of the vector potentials
$\vec{A}^\mathrm{LD}$ and $\vec{A}^\mathrm{BS}$ in the area of integration, as
is shown in the following.

First consider $E_\mathrm{ind}^\mathrm{BS}$, induced by the magnetic field of
a brain stimulation coil with current loops located close to the head. 
For simplicity a quadratic loop, shown in the left part of
Fig.~\ref{fig:sketch}, with side length~$l$ is assumed.
Thus, the vector potential~$\vec{A}^\mathrm{BS}$, and with it the electric
field $\vec{E}=-\partial_t \vec{A}^\mathrm{BS}$, forms closed loops. 
For the given loop this yields
$A_x^\mathrm{BS}(z)=A_x^\mathrm{BS}(z+l)=A_z^\mathrm{BS}(x)=A_z^\mathrm{BS}(x+l)=A^\mathrm{BS}$,
and the voltage induced in the loop can be expressed as 
$U_\mathrm{ind}^\mathrm{BS} = -\int \vec{E} \cdot \vec{dl} = \partial_t \int
\vec{A}^\mathrm{BS} \cdot \vec{dl} = 4 l \partial_t A^\mathrm{BS}$. 
This results in 
$E_\mathrm{ind}^\mathrm{BS} = - U_\mathrm{ind}^\mathrm{BS} / (4l) =
- \partial_t A^\mathrm{BS}$ 
which corresponds to eq.~\eqref{eqn:edta}.

Regarding the lightning case, a straight and vertical lightning channel
pointing in $z$-direction and an observer on perfectly conducting ground may
be assumed, so that the vector potential $\vec{A}^\mathrm{LD}$ has a
vertical component only. Consider the quadratic loop shown in the right part
of Fig.~\ref{fig:sketch}. 
Due to the small $x$-dependence of $\vec{A}^\mathrm{LD}$ we now have $A_z^\mathrm{LD}(x)\neq A_z^\mathrm{LD}(x+l)$, and the voltage induced in the loop is given by
$U_\mathrm{ind}^\mathrm{LD} = -\int \vec{E} \cdot \vec{dl} = 
\partial_t \int\vec{A}^\mathrm{LD}\cdot\vec{dl}
=  l \partial_t \left[A_z^\mathrm{LD}(x)-A_z^\mathrm{LD}(x+l)\right]
=-l^2 \partial_t \partial_x A_z^\mathrm{LD}$ 
resulting in 
$E_\mathrm{ind}^\mathrm{LD} = - U_\mathrm{ind}^\mathrm{LD} / (4l) 
= (l/4) \partial_x \partial_t A_z^\mathrm{LD}$
which is different from eq.~\eqref{eqn:edta}.

Consequently, $E_\mathrm{ind}^\mathrm{LD}$ can not be computed by
eq.~\eqref{eqn:edta} but has to be calculated from eq.~\eqref{eqn:far}. For
$E_\mathrm{ind}^\mathrm{BS}$ eqs.~\eqref{eqn:edta} and
\eqref{eqn:far} yield the same result. 
Due to the incorrect use of eq.~\eqref{eqn:edta} the results for lightning
induced electric fields obtained in Ref.~\cite{peer2010a} for average
lightning parameters are more than an order of magnitude too large (depending
on distance). Correct results for $E_\mathrm{ind}^\mathrm{LD}$ and their
consequence on the probability of cortical and retinal phosphene stimulation using
eq.~\eqref{eqn:far} are discussed in the following.

We now do not focus on one specific (average) lightning channel base current
and wave form like in Ref.~\cite{peer2010a}, but rather explore a range of
above average but still usual parameters.  
It turns out that fields induced by the previously considered long current
{\it decline phase} of return strokes (order of 100~$\mu$s) are below known
cortical phosphene thresholds for the range of considered lightning parameters
at relevant distances (i.e., more than several meters, where other lightning
effects and injury may not be expected to be dominant on an observer).  

\begin{figure}
	\includegraphics[width=0.45\textwidth, height=0.45\textwidth]{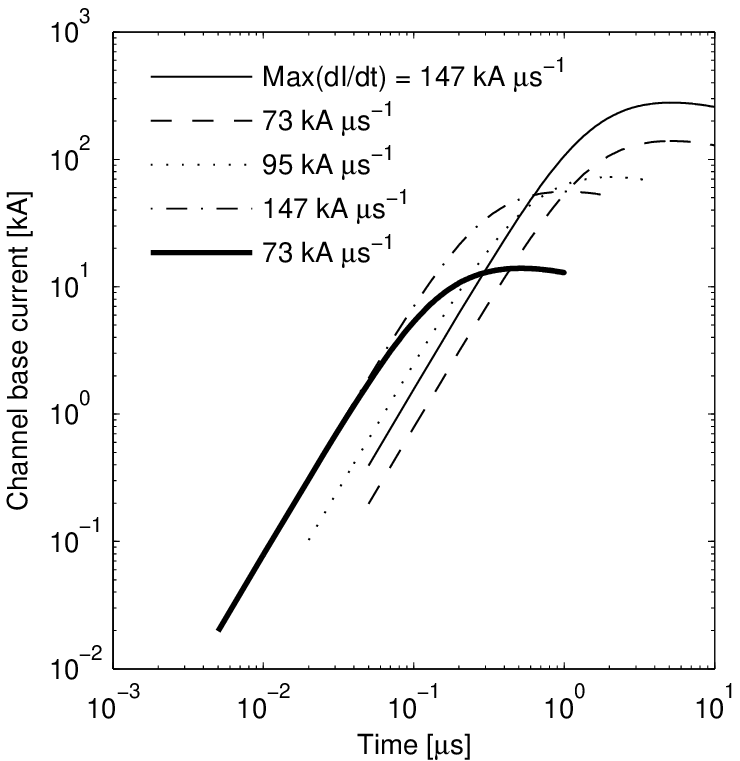}
	\includegraphics[width=0.45\textwidth, height=0.45\textwidth]{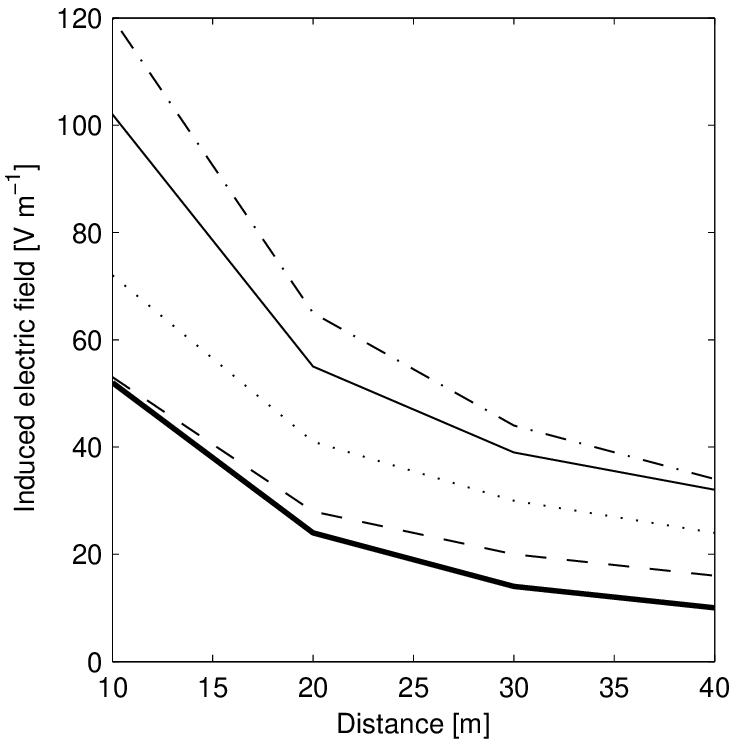}
	\caption{\sl Logarithmic plot of the initial rise phase for a range of different
          channel base current waveforms $I(t)$ (top), and associated maximum
          values of the induced electric fields  $E_\mathrm{ind}^\mathrm{LD}$
          for different distances from the lightning channel (bottom).
For the two cases with max($dI/dt$)=147~$kA/\mu s$, waveforms with
different maximum current amplitude and rise time are used.}
	\label{fig:curr}
\end{figure}

We therefore now also reconsider the possibility of an effect of the return stroke current
{\it rise phase} on cortical axon stimulation. 
Fig.~\ref{fig:curr} (top) shows the initial rise phase for different channel base
current waveforms $I(t)$ of return strokes, and Fig.~\ref{fig:curr} (bottom) shows the
corresponding associated maximum values of the induced electric fields
$E_\mathrm{ind}^\mathrm{LD}$ for different distances from the lightning
channel.
$E_\mathrm{ind}^\mathrm{LD}$ is calculated from the time varying magnetic flux
through a circular area with a cortex radius of $0.07\,\mathrm{m}$. The figure shows
that the maximum value of $E_\mathrm{ind}^\mathrm{LD}$ is mainly determined by
$\partial_t I$. 
The cortical phosphene threshold of around $20-40\,\mathrm{Vm^{-1}}$ is
exceeded in distances up to order of 50~m for return strokes
that are characterised by a current rise of 
$\frac{dI}{dt} \gtrsim 100\,\mathrm{kA\,\mu s^{-1}}$.
The duration of a single induced electric field pulse is determined by the current
rise time $(0.5-5\,\mathrm{\mu s})$ and repeated with the frequency of
multiple strokes. 


It is however not evident if these short pulses in the rise phase in the order
of microseconds actually allow stimulation of phosphenes, as no clinical
experience with similar pulse forms is available. 
In vitro experiments suggest that to fire axons may require longer
exposure ($> 100~\mu$s) to electric fields of similar strength \cite{rotem2008}.
Stimulation of cortical phosphenes by multiple lightning return strokes
therefore appears improbable for relevant parameters and distances above
several ten meters. 

On the other hand, it had already been noted in Ref.~\cite{peer2010a} that {\it
  retinal} phosphenes have a much lower threshold than their cortical
counterparts \cite{kavet2008a}, which is according to Refs.~\cite{wood2008a,litvak2002a}
in the range of $10-100\,\mathrm{mV\,m^{-1}}$.
The feasibility to stimulate retinal phosphenes with lightning induced
electric fields is therefore much higher than for cortical phosphenes. 
In Ref.~\cite{peer2010a} we expressed the point of view that when lightning
induced cortical phosphenes can be shown to possibly exist, then retinal
phosphenes are an even more likely event under the same circumstances. 
As lightning induced stimulation of cortical phosphenes has now been shown to
be much less probable, we also re-evaluate the possibility of retinal
phosphenes by means of the corrected calculations:
Indeed $E_\mathrm{ind}^\mathrm{LD}$ can reach above retinal phosphene
threshold values at distances up to order of $50\,\mathrm{m}$ from the
lightning channel also during the long return stroke decline phase of
$100-200\,\mathrm{\mu s}$ pulse duration, and in even considerably longer
distances (order of 200~m) during the short rise phase. 

Unfortunately no directly comparable specific retinal stimulation experiments could
be found in the available literature. 
While the average frequency of return strokes in a multiple lightning discharge
$(20\,\mathrm{Hz})$ coincides with the repetition frequencies usually used in
retinal stimulation  experiments, the pulse shapes of 
$E_\mathrm{ind}^\mathrm{LD}$ and $E_\mathrm{ind}^\mathrm{BS}$ considerably
differ \cite{wood2008a}: usually, the retina is stimulated by sinusoidal
waveforms with a frequency of also $20 - 45\,\mathrm{Hz}$ by TMS,
compared to return stroke pulse durations of several 100~$\mu$s. 
Studies on direct current electrical excitation of the human retina however
indeed show stimulability for short pulse durations of 250~$\mu s$ \cite{rizzo2003}.
The possibility of stimulation of retinal phosphenes by lightning
fields could of course in future be verified by physiological
investigations using comparable magnetic pulse forms.  

An experimental setup which covers both retinal and cortical stimulation
regions may indeed easily be devised with a pair of Helmholtz coils with
radius and separation larger than a human head, where currents are applied that
directly generate nonfocal magnetic fields with strengths and pulse shapes as 
calculated for realistic lightning conditions in various distances.
This suggested {\it experimentum crucis} is able to critically test the lightning
electromagnetic phosphene stimulation hypothesis.

In spite of the previous overestimation of induced electric fields in
Ref.~\cite{peer2010a}, a stimulation of phosphenes induced by lightning
electromagnetic pulses remains plausible.
The most probable site of stimulation however appears to be the retina rather
than the visual cortex.

{\bf Acknowledgment:} The original authors (JP and AK) are thankful to VC for
pointing out the problem in the previous analysis of Ref.~\cite{peer2010a}.

\end{document}